\documentclass[twocolumn,superscriptaddress,showpacs,showkeys,preprintnumbers,amsmath,amssymb,aps]{revtex4}
\usepackage{graphicx}
\usepackage{dcolumn}
\usepackage{amsmath}
\usepackage{latexsym}

\begin{document}

\def\f{\frac}
\def\rdown{\rho_{\downarrow}}
\def\pa{\partial}
\def\th{\theta}
\def\Ga{\Gamma}
\def\ka{\kappa}
\def\bea{\begin{eqnarray}}
\def\eea{\end{eqnarray}}
\def\be{\begin{equation}}
\def\ee{\end{equation}}
\def\pa{\partial}
\def\d{\delta}
\def\eps{\epsilon}
\def\th{\theta}
\def\na{\nabla}
\def\nn{\nonumber}
\def\lan{\langle}
\def\ran{\rangle}
\def\pr{\prime}
\def\rarrow{\rightarrow} 
\def\larrow{\leftarrow}

\title{Quantum current magnification in a multi-channel mesoscopic ring }

\author{Swarnali Bandopadhyay}
\email{swarnali@bose.res.in}
\author{P. Singha Deo}
\email{deo@bose.res.in}
\affiliation{S. N. Bose National Centre for Basic Sciences, JD Block,
Sector III, Salt Lake City, Kolkata 700098, India.}
\author { A. M. Jayannavar }
\email{jayan@iopb.res.in}
\affiliation{Institute of Physics, Sachivalaya Marg, Bhubaneswar 751005, India }
\date{\today}

\begin{abstract}
\noindent
We have studied the current magnification effect in a  
multi-channel open mesoscopic ring. We show that the current magnification 
effect is robust
 even in the presence of several propagating modes inspite of mode mixing and
 cancellation effects. The magnitude of circulating currents in the 
 multi-channel regime can be much larger than that in a single channel case.
 Impurities can enhance or degrade the current magnification effect depending
 sensitively on the system parameters. Circulating currents are mostly 
 associated with Fano resonances in the total transport current. We further 
 show that system-lead coupling qualitatively changes the current 
 magnification effect. 

\end{abstract}

\maketitle
\section{ Introduction }
\label{s1}
In the light of recent developments in fabrication techniques it has become
possible to make metallic or semiconductor structures having dimensions of 
a few atomic spacings. The typical size of the systems can be made smaller than the phase coherence length of the electron. Such mesoscopic systems are 
important for their possible device applications as well as their 
counterintuitive physical properties in the quantum domain 
\cite{bk-datta,bk-imry}.

 Quasi-particle current flows across an open mesoscopic ring connected to 
 electron reservoirs via
leads maintained at a constant chemical potential difference. 
The current $I$ injected by the
reservoir into one of the leads splits into $I_U$ and $I_L$ in the upper and
lower arms of the ring such that current conservation (Kirchoff's law : 
$I = I_U + I_L$) is satisfied. The size of the ring being 
smaller than the phase coherence length, the electrons in two arms in general 
will pick up different phases and their quantum mechanical superposition gives 
rise to two distinct possibilities. The first being, for some values 
of Fermi energy the currents in the two arms $I_U$ and $I_L$ are individually 
less than the total current $I$, i.e., the current in both arms flow along 
the direction of the applied field. The other possibility is that for some
values of Fermi energy, $I_U$ (or $I_L$) can be greater than
the total current $I$. In this case current conservation dictates 
$I_L$ (or $I_U$) to be negative such that $I=I_L + I_U$. The property
that the current in one of the arms is larger than the transport current is 
referred to as `{\bf current magnification}' effect 
\cite{jayan-deo,par-deo-jayan,jayan-deo-par}. 
 Magnitude of this negative current is taken to be that of 
the `circulating current'. When negative current flows in the upper arm
 the direction of the circulating current is taken to be counter clockwise
 or negative and when it flows in the lower arm its direction is taken to be 
 clockwise or positive. It should be noted that these circulating currents 
 arise in the absence of magnetic flux, but in the presence of transport 
 current (i.e., in a nonequilibrium system). When a parallel resonance circuit 
 (capacitance C connected in parallel with a combination of inductance L and
 resistance R) is driven by an external e.m.f., circulating current arises in 
 the circuit at resonance frequency \cite{shaw-bk}.
 However, the current magnification effect is absent in a circuit with two 
 parallel resistors in the presence of dc current in the classical regime. 
 In a mesoscopic ring the
intrinsic wave nature of electrons and their phase coherence gives rise to
this effect even in presence of dc driving voltage. Studies on current 
magnification effect in mesoscopic open rings have been extended to thermal 
currents \cite{mosk98} and to spin currents in presence of Aharonov-Casher 
flux \cite{choi98}. This effect has been studied in the presence of a
spin-flip scatterer which causes dephasing of electronic motion
\cite{joshi01,colin1}.  

The predicted magnitude of the circulating current densities can indeed be 
very large \cite{par-deo-jayan} and has been termed as `giant persistent 
currents' \cite{yi,wu}. Recently the current magnification effect has been 
shown to occur in mesoscopic hybrid system at equilibrium in the presence of a 
magnetic flux and in the absence of transport current \cite{colin2,colin3}.
So far all studies on current magnification have been restricted to the case 
of one dimensional (single channel) systems only. In this
work we go beyond the single channel regime to a multi-channel one. 
Multi-channel systems are a closer realisation to the experimental 
systems \cite{cerni97,cond03}
due to their finite width in the transverse direction of propagation of 
currents. In the present work we show that inspite of the contributions from 
large number of 
different modes and mode-mixing, current magnification sustains for various 
length ratios of the two arms of the ring. The connection between current 
magnification and Fano resonance in the total current is shown. In very
special cases current magnification is shown to occur near Briet-Wigner
type resonances. The strong 
qualitative dependence of current magnification and system-reservoir coupling 
strength is also established. 

\section{ Description of the System }
\label{s2}
\begin{figure}[t]
\includegraphics [width=8cm]{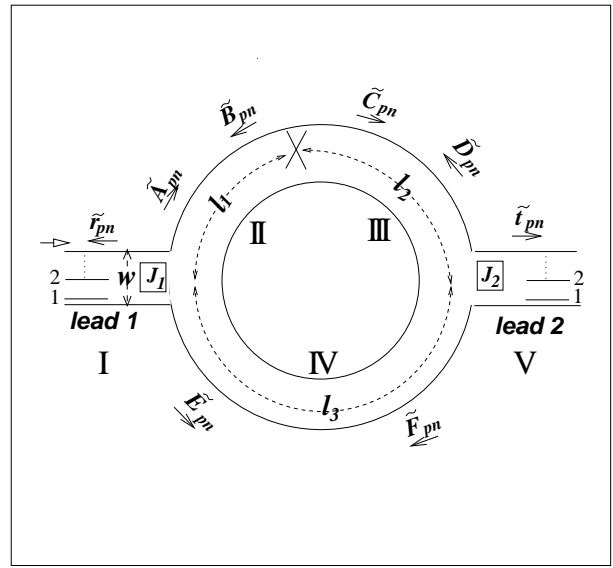}
\caption{Schematic diagram of an open multi-channel mesoscopic ring of perimeter $L=l_1+l_2+l_3$ connected through the leads $1$ and $2$ to electron reservoirs (not shown in figure). Both the ring and the leads have the same width $W$. Several transverse modes are shown by horizontal lines in the leads. A delta function type static scatterer $V_d(x,y) = V \d(x-l_1) \d(y-y_i) $ of strength $V$ is shown in the upper arm at $\times$. $\eps$ denotes the 
coupling strength between leads and the ring. 
\label{system}
}
\end{figure}

In our present work we consider a quasi-one-dimensional (Q1D) ring of 
perimeter $L$ and width $W$ with 
$L >> W$ as shown in Fig. \ref{system}. The two leads that connect this 
system to the electron reservoirs have the same width as that of the ring. 
The length of the lower arm of the ring is $l_3$ while that of the 
upper arm is $l_1+l_2$. An impurity $\d$ function potential 
$V\d(x-l_1)\d(y-y_i)$ is embedded in the upper arm. 
The electrons can propagate freely along the length of 
the ring and leads but their motion is confined along the transverse direction. We consider hard wall confinement potential 
along the transverse direction. Due to this confinement, infinite number of 
transverse modes are generated in the system. If the energy of the electrons 
is such that the corresponding wave number is real then the mode is termed as 
propagating, on the other hand, if the wave number is imaginary it is termed as 
evanescent. The widths of the ring and leads being equal the number of 
propagating and evanescent modes are same in these two.

We consider the non-interacting electrons in the system. The system size is 
taken to be smaller than the phase coherence length $l_\phi$ and the phase 
randomizing inelastic scattering is considered only inside the reservoir. 
Scattering inside the system maintains the phase coherence. This necessitates
only static scatterers in the system which in our case are a delta-function 
potential and junction scatterers at $J1$ and $J2$. We neglect all phase 
randomizing scattering like 
electron-phonon interaction inside the system. The left reservoir ($R_L$) and 
the right reservoir ($R_R$) have chemical potentials $\mu_L$ and $\mu_R$ 
respectively.  
When $\mu_L > \mu_R$ current flows from $R_L$ to $R_R$ and vice-versa. We are 
interested in the linear response regime where currents are related to the 
transmission across the system at the Fermi energy (Landauer-B\"uttiker formula, \cite{bk-datta}). We consider 
that the electrons enter the system through the left lead and come out 
through the right lead. Due to mirror symmetry, results remain the same 
if the flow of electrons is reversed with the direction of circulating 
current getting reversed. This ensures absence of circulating current in 
equilibrium at zero magnetic field. Magnetic field breaks the time reversal 
symmetry and hence one can obtain persistent currents circulating across the 
ring in equilibrium \cite{but-im-lan}. These currents have been observed 
experimentally both in open and closed systems \cite{levy,bk-imry}.

For no loss of generality we have considered the situation wherein no 
mode mixing between 
different transverse modes occur at the junctions. The ring and the lead are
connected via junction scattering matrices at $J1$ and $J2$. The junction
scattering matrices are same for both the junctions $J1$ and $J2$. The coupling
between either sides of the junction for the modes with same transverse 
quantum number is given by \cite{butt85}
\be
S_J = \left(\begin{array}{ccc}
-(a + b ) & \sqrt{\eps} & \sqrt{\eps}\\
\sqrt{\eps} & a & b\\
\sqrt{\eps} & b & a
\end{array}\right)
\label{eq:S-J}
\ee
where $a = \frac{1}{2} \left( \sqrt{1-2\eps} - 1 \right)$ and 
$b = \frac{1}{2} \left( \sqrt{1-2\eps} + 1 \right)$. $\eps$ is
a coupling parameter with values $ 0 \le \eps \le 0.5$. When
$\eps \to 0$ the system and the reservoir are decoupled while for 
$\eps \to 0.5$ these two are strongly coupled. This $S$-matrix
satisfies the conservation of current \cite{shapiro}.
The above $S$-matrix is independent of the incident energy and the index of the 
transverse modes. The presence of the elastic scatterer, namely, $\delta$
function potential in the upper arm mixes different propagating and evanescent 
modes along with extra phase shift in the same mode. 

When electrons are injected in the $p$-th propagating mode, the total
wave function in the left lead (region I) is given by
\be
\psi\big |_{\mbox {I}} = \sqrt{\it N}  e^{ik_px} \chi_p(y) +  \sum_n 
r_{pn} e^{-ik_nx} \chi_n(y) \mbox{  ,}
\label{inwv}
\ee
 where $k_p$ is the longitudinal wavevector corresponding to $p$-th mode along 
 the direction of propagation. 
  Here $r_{pn}$ describes reflection amplitude from $p$-th mode to $n$-th
mode, $\chi_n(y) $ represents the $n$th transverse mode where $y$ is
the coordinate along the transverse direction and $\sum_n$
denotes summation over $n$ including $p$.
The normalisation factor $\sqrt{\it N}$ is determined by noting that 
the current density injected by the reservoir in a small energy interval $dE$ 
in the $p$-th propagating mode is 
\be
dj_{p_{in}} =  e v_p \frac{dn_p}{dE} f(E) dE 
\label{incur}
\ee 
where $f(E)$ is the Fermi distribution function, $\frac{dn_p}{dE} = 
\frac{2}{hv_p}$ is the density of states (DOS) 
in the perfect wire and $v_p = \frac{\hbar k_p}{m_e}$. For our 
zero temperature calculations $f(E) = 1$ for occupied states. 
The wave function $\psi_p\big |_{\mbox {I}}$ gives the  
incident current density $ dj_{p_{in}} = 
\frac{2e}{h} dE $, which in turn is independent of the propagating mode in 
which the electron is incident if $ N = \frac{2e}{hv_p} dE $. Here $dE$ 
denotes an energy interval around Fermi energy and 
hence change in incident energy would mean a change in the Fermi 
energy of electrons eminating from the reservoirs. 

The wave functions in all other regions are
\bea
\psi\big |_{\mbox {II}} &=& \sum_n \big(A_{pn} e^{ik_nx} +  
B_{pn} e^{-ik_nx}\big) \chi_n(y)
\label{wv2}\\
\psi\big |_{\mbox {III}} &=& \sum_n \big(C_{pn} e^{ik_nx} +  D_{pn} e^{-ik_nx}
\big) \chi_n(y)\label{wv3}\\
\psi\big |_{\mbox {IV}} &=& \sum_n \big(E_{pn} e^{ik_nx} +  F_{pn} e^{-ik_nx}
\big) \chi_n(y)\label{wv4}\\
\psi\big |_{\mbox {V}} &=& \sum_n t_{pn} e^{ik_nx} \chi_n(y) 
\label{wv5}
\eea
where $n$ stands for all available propagating modes including $p$.  

$S_J$ connects the incoming and outgoing amplitudes of the $p$-th mode 
at $J_1$ via
\be
\left(\begin{array}{c} \tilde r_{pp} \\ \tilde A_{pp} \\ \tilde E_{pp}
\end{array}\right)
 = S_J \left(\begin{array}{c} 1 \\ \tilde B_{pp}\\ \tilde F_{pp}
\end{array}\right)
\label{eq:lead1-cal}
\ee
where any new amplitude $\tilde A_{pn} $ is connected to its earlier definition
$ A_{pn} $ by
$$ \tilde A_{pn} = \sqrt{v_p} \sqrt{\frac{h}{2e}} (\sqrt{dE})^{-1} A_{pn}$$ 
In further calculations all the tilded amplitudes carry the same
meaning as above.
$S_J$ connects the incoming and outgoing amplitudes of all the other 
propagating modes $m$ ($m\ne p$ ) at $J_1$ via
\be
\left(\begin{array}{c} \tilde r_{pm} \\ \tilde A_{pm} \\ \tilde E_{pm}
\end{array}\right)
 = S_J \left(\begin{array}{c} 0 \\ \tilde B_{pm} \\ \tilde F_{pm}
\end{array}\right)
\label{eq:lead1-cal}
\ee
Similarly, the incoming amplitudes (0,$\tilde C_{pn},\tilde E_{pn} $) and 
outgoing amplitudes ($\tilde t_{pn},\tilde D_{pn},\tilde F_{pn}$) at the 
junction $J_2$ are connected via the same scattering matrix $S_J$.

The elastic scattering at the impurity is described by 
\be
\left(\begin{array}{c} \tilde B_{p1} \, e^{-i k_1 l_1} \\ 
\tilde B_{p2} \, e^{-i k_2 l_1} \\  \cdots \\ \cdots \\ \tilde B_{pP}
 \, e^{-i k_{p_P} l_1}\\ \tilde C_{p1} \\ \tilde C_{p2} \\ \cdots \\ 
\cdots \\\tilde C_{pP}  \end{array}\right)
 = \tilde S \,\, \left(\begin{array}{c} \tilde A_{p1} \, e^{i k_1 l_1} \\ 
\tilde A_{p2} \, e^{i k_2 l_1} \\  \cdots \\ \cdots \\ 
\tilde A_{pP} \, e^{i k_P l_1}\\
\tilde D_{p1} \\ \tilde D_{p2} \\ \cdots \\ \cdots \\\tilde D_{pP}
\end{array}\right)
\label{eq:imp}
\ee
$$ \mbox{ where   } \tilde S = \left(\begin{array}{cc} 
\tilde R & \tilde T \\ \tilde T & \tilde R \end{array}\right)
\label{Simp}$$ 
and both $\tilde R$ and $\tilde T$ are matrices of order $P \times P$, 
$P$ being the maximum number of propagating modes in the system depending
on a given Fermi energy. Here
$$\tilde R = \left(\begin{array}{ccccc}\tilde \rho_{11} & \tilde \rho_{12} & \cdots 
& \cdots & \tilde \rho_{1P} \\ 
\tilde \rho_{21} & \tilde \rho_{22} & \cdots & \cdots & \tilde \rho_{2P}\\
\cdots & \cdots & \cdots  & \cdots & \cdots \\
\cdots & \cdots & \cdots  & \cdots & \cdots \\
\tilde \rho_{1P} & \tilde \rho_{2P} & \cdots & 
\cdots & \tilde \rho_{PP} 
\end{array}\right)
$$
and
$$\tilde T = \left(\begin{array}{ccccc}\tilde \tau_{11} & \tilde \tau_{12} & \cdots 
& \cdots & \tilde \tau_{1P} \\ 
\tilde \tau_{21} & \tilde \tau_{22} & \cdots & \cdots & \tilde \tau_{2P}\\
\cdots & \cdots & \cdots  & \cdots & \cdots \\
\cdots & \cdots & \cdots  & \cdots & \cdots \\
\tilde \tau_{1P} & \tilde \tau_{2P} & \cdots & 
\cdots & \tilde \tau_{PP} 
\end{array}\right) \mbox{       ,}
$$
$$ \mbox{where   } \tilde \rho_{mn} = \frac{-i\frac{\Ga_{mn}}{2\sqrt{k_mk_n}}}
{1+\sum_j^e\frac{\Ga_{jj}}{2\ka_j}+i\sum_j^p\frac{\Ga_{jj}}{2k_j}}\,\, .$$
$\sum^e$ represents sum over all the evanescent modes
and $\sum^p$ represents sum over all the propagating modes
and the modes $m$ and $n$ are assumed as propagating. The intermode 
(i.e. $m \ne n$)
transmission amplitudes are $\tilde \tau_{mn} = \tilde \rho_{mn}$ and intramode
transmission amplitudes are $\tilde \tau_{nn} = 1+\tilde \rho_{nn}$. For 
details see Ref.\cite{bag90}. $\Ga_{mn}$ can be calculated using
$$ \Ga_{mn} = \frac{2m_eV}{\hbar^2} \, \chi_n^*(y_i) \chi_m(y_i)   .$$
In Eq.~\ref{eq:imp}, while writing $\tilde A_{pn} $, $\tilde B_{pn}$ the 
origin is taken to be at the junction $J_1$ whereas in writing 
$\tilde C_{pn}$, $\tilde D_{pn}$ the origin is taken at the scatterer.
For details of the $S$-matrix elements for a multi-channel scattering problem
see Ref.~\cite{bag90}. Note that different elements of the $S$-matrix contains 
information about the propagating modes as well as all the infinite number of
evanescent modes arising out of transverse confinement \cite{bag90}. 

For any given incident electron in the $p$-th mode in the lead with energy $E$ 
the current in the $n$-th mode in region II is given by
\bea
dj_{p,n_{_{LU}}} &=& v_n \,( \,|A_{pn}|^2 - |B_{pn}|^2 \,)\nn\\
&=& (\, |\tilde A_{pn}|^2 - |\tilde B_{pn}|^2\, ) \,\frac{2e}{h}
\label{cur:LU}
\eea
Currents in all other portions of the ring can be calculated similarly. The
partial current densities $dj_{{p,n}_{LU}}$ are obtained after integrating 
the local currents along the transverse $y$ direction. 
If $dj_{p,n}$ is the current density in the $n$-th propagating mode in any 
segment of the system then the total current in that segment is given by
\be
dj = \sum_{p=1}^{P} \,dj_p(s) = \sum_{p=1}^{P} \,\,\sum_{n=1}^{P} \, dj_{p,n}
\label{current}
\ee
where `$p$' denotes the propagating mode in which the electrons are injected
from the reservoir.

We use scattering matrices at the two junctions and at the scatterer site, 
$\times$, to calculate all the amplitudes and then find out the total 
current density ($dj_T$), 
the current density in the upper arm ($dj_U$) as well as in the 
lower arm ($dj_L$). Thus
\bea
dj_T & = &\sum_{p=1}^P \sum_{n=1}^P \big|\tilde t_{pn}\big|^2 
\frac{2e}{h} \nn \\ 
& = & \sum_{p=1}^P \big(1-\sum_{n=1}^P  \big|\tilde r_{pn}\big|^2 \big) 
\frac{2e}{h}   \label{totcurexpr}
\eea
We study these currents as a function of the incident electron energies.

\section{results and discussions}
\label{S3}
\begin{figure*}[t]
\begin{center}
\includegraphics [height=10cm,width=16cm]{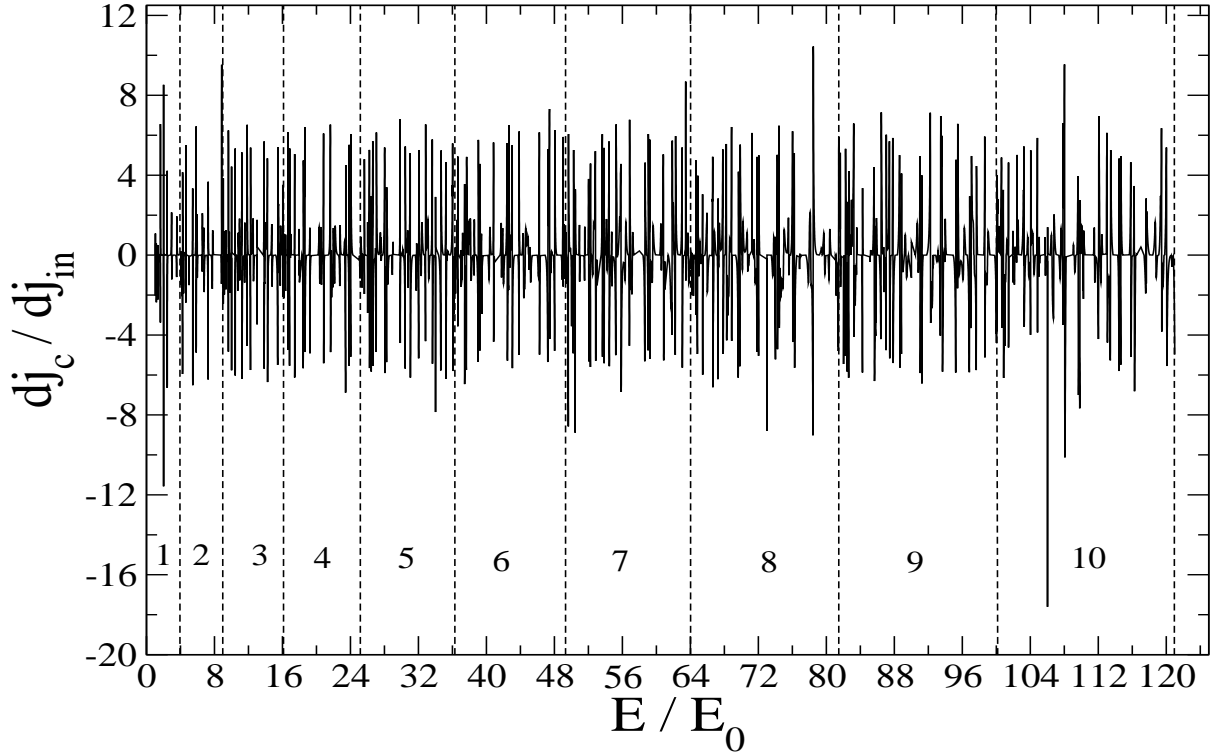}
\end{center}
\caption{Plot of circulating current density $dj_c/dj_{in}$ in the ring as a function of $E/E_0$ of the electron. In the whole energy range, we have 1 to 10 propagating modes (corresponding energy-range are indicated by vertical dashed lines). The different system parameters are $l_1=3.5$, $l_2=2.5$, $l_3=4.0$, $W=1$, $V=1$, $y_i=0.21W$, $\eps=0.2$. 
\label{result-weak1}
}
\end{figure*}

The circulating current density $dj_c$ is the magnitude of the negative part 
of $dj_U$ or $dj_L$ as mentioned earlier. When $dj_U$ is negative the direction 
of circulating electron current is anticlockwise (negative) and when $dj_L$ is 
negative then it is clockwise (positive).  
A circulating current in a loop 
gives rise to an orbital magnetic moment (Ampere's law). By our convention, 
positive $dj_c$ indicates an `up' magnetic moment whereas negative $dj_c$ 
indicates a `down' one. We plot all current densities in units of incident 
current density $dj_{in}=\f{2e}{h}dE$ and all energies in units of ground state 
energy of the lowest transverse mode, $E_0=\f{\pi^2\hbar^2}{2m_eW^2}$. 
In all our calculations we have considered 
$500$ evanescent modes. 
Increasing the strength of the impurity potential causes the coupling of 
higher number of evanescent modes and hence for large impurity 
potential strengths one need to incorporate larger number of evanescent modes. 
\begin{figure*}[t]
\begin{center}
\includegraphics [height=10cm,width=16cm]{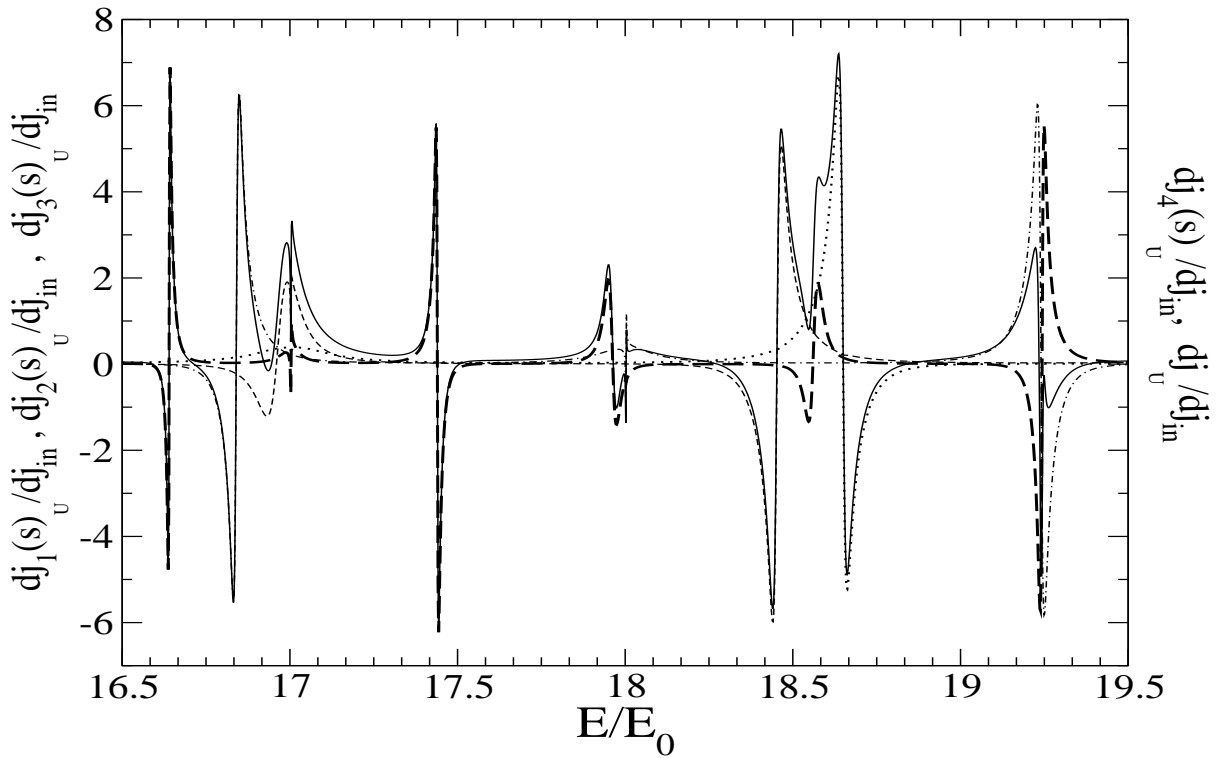}
\end{center}
\caption{Plots of different partial current densities and the total current in the upper arm of the ring as a function of $E/E_0$. The dotted curve gives $dj_1(s)_{_U}/dj_{in}$, the dashed curve gives $dj_2(s)_{_U}/dj_{in}$, the dash-dotted curve gives $dj_3(s)_{_U}/dj_{in}$, the long-dashed one is for $dj_4(s)_{_U}/dj_{in}$ and the solid one is for $dj_U/dj_{in}$. In the above energy range we have 4 propagating modes. The different system parameters are $l_1=3.5$, $l_2=2.5$, $l_3=4.0$, $W=1$, $V=1$, $y_i=0.21W$, $\eps=0.2$. 
\label{weak2-mix}
}
\end{figure*}
\begin{figure*}[t]
\begin{center}
\includegraphics [height=10cm,width=16cm]{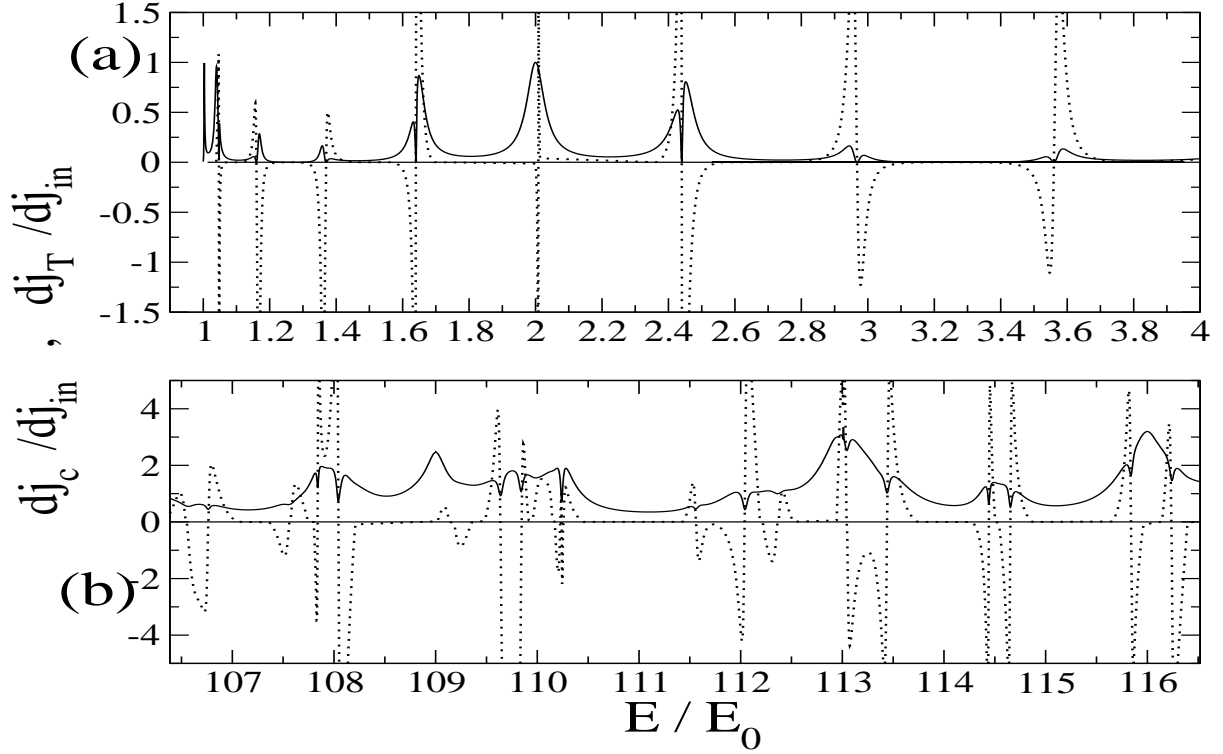}
\end{center}
\caption{Plot of the circulating current $dj_c/dj_{in}$ (dotted lines) and the total current $dj_T/dj_{in}$ (solid lines). Both the functions are plotted versus $E/E_0$. The different system parameters are $l_1=3.5$, $l_2=2.5$, $l_3=4.0$, $W=1$, $V=1$, $y_i=0.21W$, $\eps=0.2$. 
\label{result-weak2}
}
\end{figure*}
\begin{figure}[hbp!]
\begin{center}
\includegraphics [width=8cm]{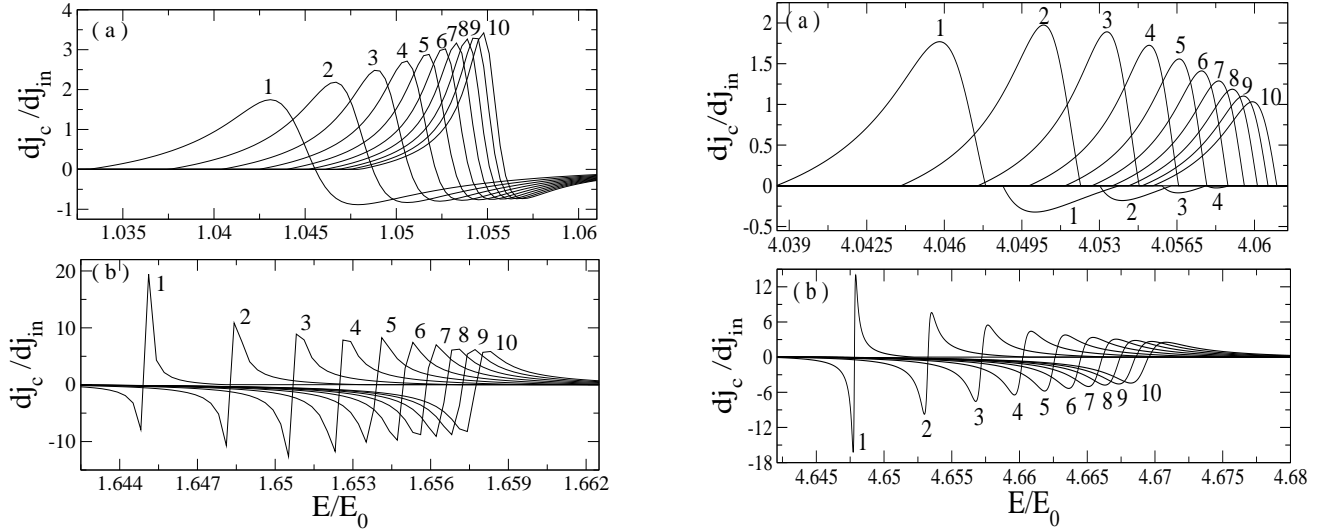}
\end{center}
\caption{ Plots of circulating current $dj_c/dj_{in}$ versus incident energy $E/E_0$ in the range of single mode propagation. Curves $1,2,3,4,\cdots,10$ are for potential strengths  $V=0.5, 1.0, 1.5, 2.0,\cdots, 5.0$ respectively. The other system parameters are $l_1=3.125$, $l_2=3.125$, $l_3=3.75$, $W=1$, $y_i=0.21W$, $\eps=4/9$. 
\label{1ch-difV}
}
\end{figure}
\begin{figure}[hbp!]
\includegraphics [height=7cm, width=8cm]{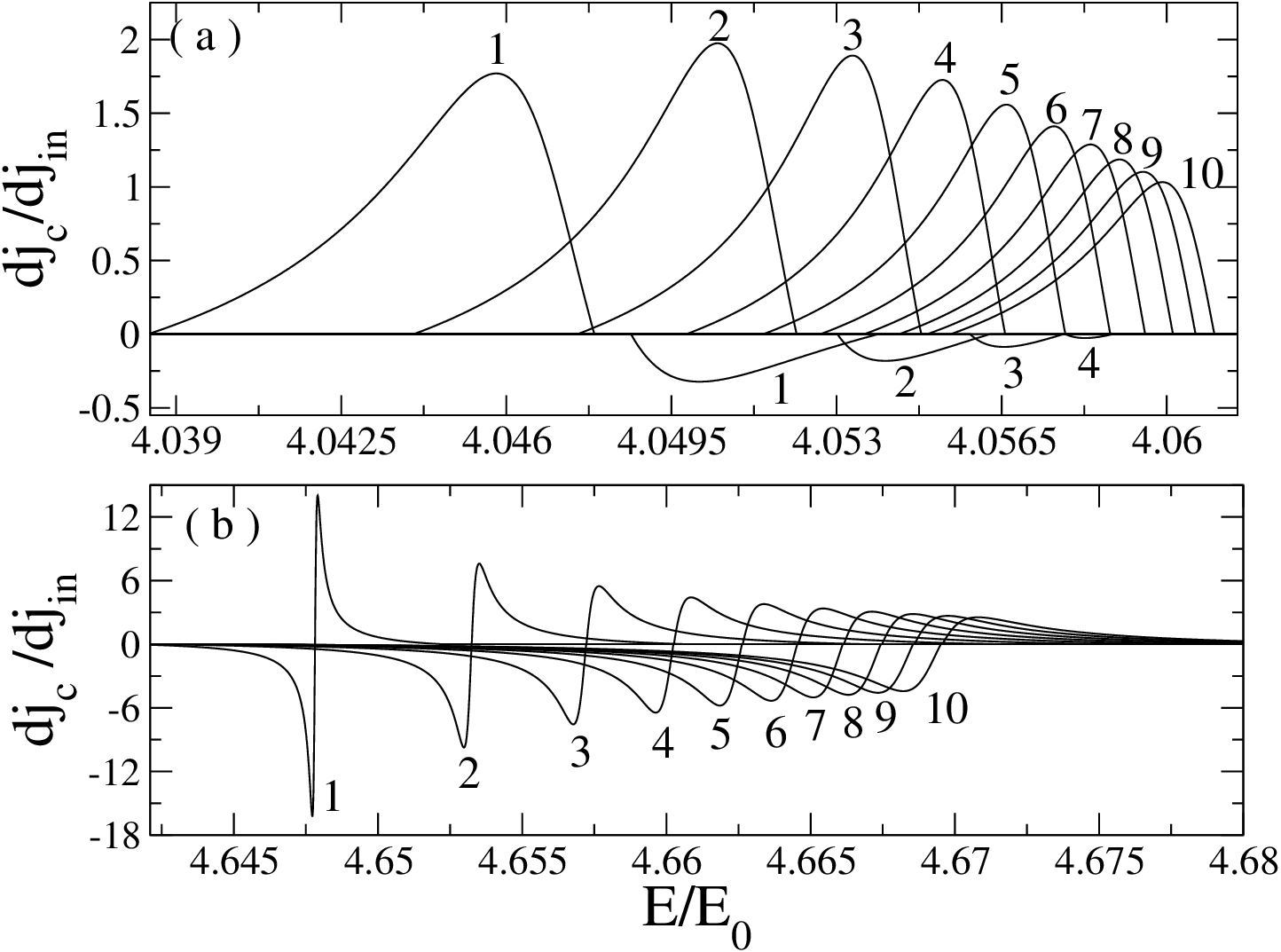}
\caption{ Plots of circulating current $dj_c/dj_{in}$ versus incident energy $E/E_0$ in the range of two mode propagation. Curves $1,2,3,4,\cdots,10$ represent potential strengths  $V=0.5, 1.0, 1.5, 2.0,\cdots, 5.0$ respectively. The other system parameters are $l_1=3.125$, $l_2=3.125$, $l_3=3.75$, $W=1$,$y_i=0.21W$, $\eps=4/9$. 
\label{2ch-difV}
}
\end{figure}

We first study the case for which the system is weakly coupled with the 
leads (Fig.~\ref{result-weak1}). This coupling can be controlled by 
appropriately changing the values of $\eps$. All the physical parameters 
are indicated in the figure caption. The upper and lower arms of the ring 
have different lengths. From the plot of circulating current density vs. 
energy (Fig.~\ref{result-weak1}) we observe that there is current magnification 
of almost same magnitude with similar frequency of occurrence over the entire 
energy range. The total number of propagating modes in the lead incident 
on the ring vary throughout this 
energy scale from one to ten as indicated in Fig.~\ref{result-weak1}.
Number of propagating modes in the lead and the ring are same. Between
different propagating modes there are several resonances around which
current magnification takes place \cite{jayan-deo,par-deo-jayan}. These
resonances approximately occur around 
$E_r = \hbar^2  \big(\frac{2r\pi}{L}\big)^2 $ , where
$E_r $ is the energy eigenvalues of the isolated ring of length $L$. The
small deviations of resonances from these values is due to multi-channel
nature of our problem along with impurity potential which causes mode mixing.
When there are say ten propagating modes, to obtain total current in the upper 
arm we have to add hundred values of partial currents [Eq.~\ref{current}] due 
to different modes. Though individual partial current density show oscillatory 
behaviour the magnitude of the total circulating current remains of 
the same order
when there is only one propagating mode in the system. This can be explicitly
seen in Fig.~\ref{result-weak1} throughout the energy range with one to ten 
propagating modes. This clearly indicates that current magnification effect is 
robust even in multi-channel systems inspite of contributions from several 
propagating modes and mode mixing. To see the mode mixing and
the cancellation effects we have considered the case 
where there are four propagating modes in Fig.~\ref{weak2-mix}. Hence to obtain 
total current in the
upper arm we have to calculate sixteen partial currents [Eq.\ref{current}].
In Fig. \ref{weak2-mix}, for simplicity instead of considering sixteen partial
currents we have plotted four values of current densities $dj_1(s)_{_U},
dj_2(s)_{_U}, dj_3(s)_{_U}, dj_4(s)_{_U}$ and total current $dj_U$. 
Here $ dj_i(s)_{_U} = \sum_{n=1}^4 dj_{i,n} \, , i=1,2,3,4 $. $dj_{i,n}$ is sum 
over partial current densities in the four propagating modes in the upper 
arm when 
electron is incident in the $i$-th propagating mode. Total current in the 
upper arm is given by 
$ dj_U = dj_1(s)_{_U}+ dj_2(s)_{_U}+ dj_3(s)_{_U}+ dj_4(s)_{_U}$. 
Negative currents in this graph represents the existence of circulating 
current in partial current
densities. Each $dj_i(s)_{_U}$ show oscillatory and complex pattern. 
The total current $ dj_U$ still exhibits negative part (current magnification) 
inspite of cancellation effects arising due to mode mixing.

\begin{figure*}[t]
\includegraphics [height=7cm,width=16cm]{swarnaliFig7.eps}
\caption{ The circulating current $dj_c/dj_{in}$ versus $E/E_0$ is plotted for repulsive potential $V=1$ in strong coupling regime $\eps=0.48$. The different system parameters are $l_1=3.5$, $l_2=2.5$, $l_3=4.0$, $W=1$, $y_i=0.21W$. 
\label{result-repul-strong}
}
\vskip2cm
\includegraphics [height=8cm,width=16cm]{swarnaliFig8.eps}
\caption{In $(a)$ the partial current $dj_1(s)_{_L}/dj_{in}$ and in $(b)$ the partial current $dj_2(s)_{_L}/dj_{in}$  in the lower arm of the ring are plotted as a function of $E/E_0$. In both $(a)$ and $(b)$ dashed curves are for $\eps=0.2$ and the solid curves are for $\eps=0.48$. The other system parameters are $l_1=3.5$, $l_2=2.5$, $l_3=4.0$, $W=1$, $V=1$, $y_i=0.21W$. 
\label{fig:compar-modes}
}
\end{figure*}
\begin{figure}[hbp!]
\includegraphics[width=8cm]{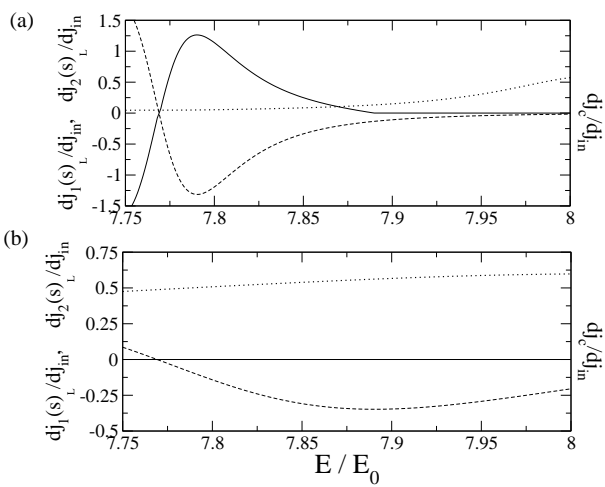}
\caption{ Both in $(a)$ and $(b)$, the dashed curve gives the partial current $dj_1(s)_{_L}/dj_{in}$, the dotted curve gives the partial current $dj_2(s)_{_L}/dj_{in}$ in the lower arm of the ring and the solid curve gives the circulating current $dj_c/dj_{in}$ in the ring. All three functions are plotted as a function of incident energy $E/E_0$ of the electron. $(a)$ is for weak coupling $\eps = 0.2$ while $(b)$ is for strong coupling $\eps = 0.48$. Other system parameters are $l_1=3.5$, $l_2=2.5$, $l_3=4.0$, $W=1$, $V=1$, $y_i=0.21W$. 
\label{fig:hiloeps}
}
\end{figure}
\begin{figure*}[t]
\includegraphics[height=8cm,width=16cm]{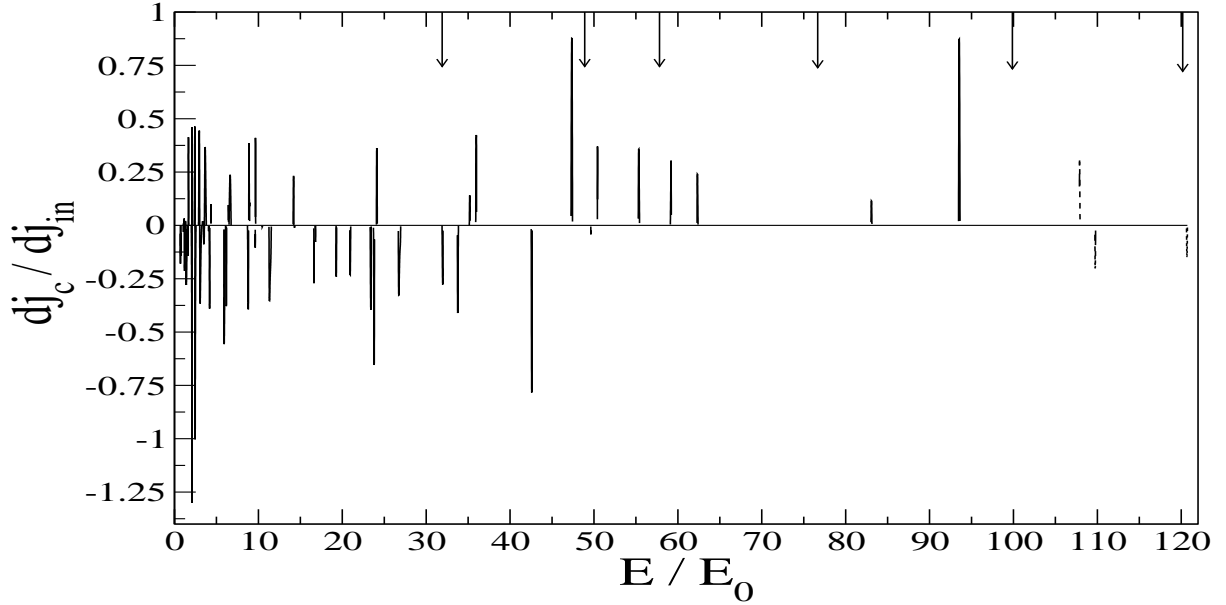}
\caption{ The circulating current $dj_c/dj_{in}$ vs. $E/E_0$ is plotted for strong coupling $\eps=0.48$ and attractive potential $V=-2.5$. The arrows on the graph denotes the positions of different quasi-bound-states in the available energy range, $31.87\,E_0, 49\,E_0, 57.7\,E_0, 76.53\,E_0, 100\,E_0, 120.12\,E_0$. The other system parameters are $l_1=3.5$, $l_2=2.5$, $l_3=4.0$, $W=1$, $y_i=0.21W$. 
\label{result-attr-strong}
}
\end{figure*}
\begin{figure}[hbp!]
\includegraphics[width=8cm]{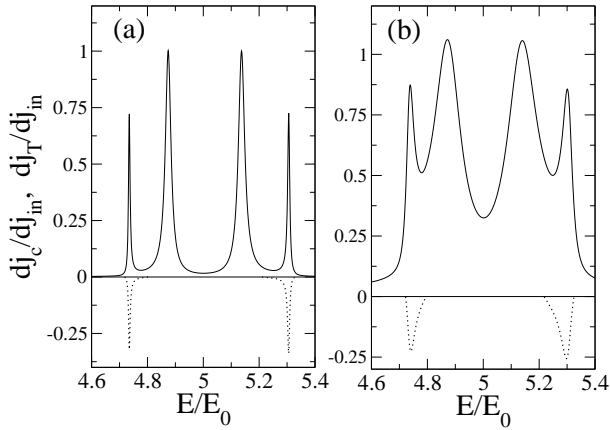}
\caption{$dj_T/dj_{in}$ (solid curve) and $dj_c/dj_{in}$ (dotted curve) are plotted as a function of $E/E_0$ for coupling strengths($\eps)$ 0.05 ((a)) and 0.2 ((b)) in presence of two propagating modes. Other system parameters are $V=-2.4$, $l_1=2.5$, $l_2=1.5$, $l_3=3.0$, $W=1$, $y_i=0.21W$. The quasi-bound-state is at $5.1025 E_0$.  
\label{asym-nearbnd1}
}
\end{figure}

To see in detail the nature of current magnification vis-a-vis total transport 
current in lead we consider a case where there is only one propagating mode
(Fig. \ref{result-weak2}(a)) and separately another case wherein number of 
propagating modes are ten (Fig. \ref{result-weak2}(b)). In these figures we 
have plotted the total transport current and circulating currents as a function of Fermi energy with the other parameters are mentioned in figure caption.
We see a current magnification whenever there is a partial minimum in the total 
current that flows through the system which in turn is measured at the leads. 
This is consistent with earlier observations seen in the case of one 
dimensional system \cite{jayan-deo}.

When only one channel is propagating the total 
current is proportional to the transmission coefficient \cite{bk-datta}. 
A closer look at these minima shows that we obtain current magnification of 
either sign around every maxima-minima pair in total current. In 
Ref.~\cite{jayan-deo,yi,wu} the current magnification of a pure 1D quantum 
ring having no impurity has been related to Fano resonance 
(asymmetric zero-pole structure) in the transmission coefficient. 
In multi-channel transmission Fano 
zero-pole line shape gets replaced by an asymmetric maximum-minimum lineshape
\cite{swarnali1}. We found this Fano type asymmetric maxima-minima lineshape
at each energy point of current magnification shown in Fig. \ref{result-weak1}.
From a first look in the range $ 1.8E_0 < E < 2.2E_0$ in 
Fig. \ref{result-weak2}(a) it appears that we have 
current magnification near a total-current maximum. But a closer scan reveals 
that there is indeed a very sharp Fano-type asymmetric maxima-minima lineshape
at this point, though it is not visible in the graph. At current 
magnification the presence of a quasi-bound state of circulating current in 
the ring gives rise to this Fano-type lineshape to the total current.
The circulating current changes sign more sharply and shows stronger current 
magnification where Fano lineshape is sharper and narrower. This feature is 
somewhat equivalent to the classical parallel LCR resonance circuit in which 
the higher $Q$-values indicate higher current magnification and sharper 
minimum at resonant frequencies.
These features remain intact for whole energy scale even if there are more than one propagating modes contributing (see Fig. \ref{result-weak2}(b)).

We observe that for a symmetric ring ($l_1+l_2=l_3$) frequency of
the occurence of
current magnification reduces throughout this energy scale considered earlier. 
This is understandable as in the absence of any impurity and magnetic field 
an asymmetric 1D ring shows current magnification ~\cite{jayan-deo}, meaning 
that asymmetry in length ratios of a ring favours this effect.
Pareek {\it et al.}~\cite{par-deo-jayan} have shown that for
a 1D ring one can have regions of incident energies where current magnification gets enhanced with the increase in the impurity potential strength. We 
investigate this effect for the case of our multi-channel ring.
In order to compare with Ref.~\cite{par-deo-jayan} we have calculated the 
effects of potential using Griffiths boundary condition or coupling parameter 
$\eps=4/9$ at the junctions.
In Fig.~\ref{1ch-difV} and Fig.~\ref{2ch-difV} we have shown the 
variation of the current magnification for two different peaks
in the appropriate energy ranges. The Fig.~\ref{1ch-difV} is for
single channel case while Fig.~\ref{2ch-difV} is for two channel.
From Fig.~\ref{1ch-difV}(a) and Fig.~\ref{2ch-difV}(a) we notice that 
the current magnification effect get enhanced with the increase in  
strength of the impurity potential while the opposite is true
for the case considered in Fig.~\ref{2ch-difV}(b). A closer look
at Fig.~\ref{1ch-difV}(b) reveals that peak in the negative part of the
circulating current density first increases and then decreases as we vary 
continuously the strength of the impurity potential. Thus impurities in 
the system can either enhance or decrease the current magnification
effect. The enhancement of the circulating current densities is a 
counter-intuitive effect, in the light of the fact that impurity
generally degrades the transport current in the system.

B{\"u}ttiker~\cite{butt85} has shown that when
the ring is threaded by a magnetic flux $\Phi$, as the coupling goes towards 
the strong coupling regime 
($\eps\rightarrow 0.5$) the amplitude of persistent current reduces due to 
increased dephasing. This is a quantitative change in 
persistent current due to broadening of energy levels with increasing
coupling strength. To examine the effect of system-reservoir coupling strength 
on current magnification in multi-channel ring in absence of $\Phi$, 
we have calculated 
circulating current for $\eps=0.48$. We observe that the frequency of 
current magnification as well as the magnitude of circulating currents reduce 
significantly in the whole energy range (Fig. \ref{result-repul-strong}) 
compared to that observed in Fig. \ref{result-weak1}. 
This indicates that in Q1D coupling strength alters the nature of current 
magnification effect in a non-trivial manner.
The total current magnification in the ring is due to a 
summation over current 
magnifications corresponding to each incident mode allowed for a given 
incident energy. As we increase the coupling strength $\eps$, we observe 
that the contribution to current magnification from electrons injected in 
each incident mode goes from a narrower  and stronger to a broader and 
weaker shape with respect to the corresponding energies
(Fig~\ref{fig:compar-modes}). The more broader they get, the more 
cancellation of current magnification occurs due to overlap of different 
incident modes. In Fig.~\ref{fig:hiloeps} we have shown contributions from 
first and second incident modes in the upper arm of the ring when only 
two modes are propagating. In this energy range we observe current 
magnification for $\eps=0.2$ (upper graph) and no current magnification 
for $\eps=0.48$ (lower graph). From lower graph ($\eps=0.48$) it is evident 
that though the contribution of current due to incident mode-$1$ (dashed line)
is negative and thus should give rise to current magnification, 
the contribution from incident mode-$2$ (dotted line) cancels it off and 
we observe no net current magnification in this energy range. From
the upper graph ($\eps=0.2$) it is clear that in the energy range where 
contribution of current due to incident mode-$1$ (dashed line) is negative, 
the contribution from incident mode-$2$ (dotted line) is almost zero as the 
contribution to current magnification from each incident mode is very sharp 
for low $\eps$-values.
 Hence we obtain a net current magnification in this energy range for $\eps=0.2$ though current magnification is absent for $\eps=0.48$ in the same 
 energy range. 
 Thus system-reservoir coupling strength alters current magnification effect 
 in a multi-channel mesoscopic ring not only quantitatively, but it also has a 
 strong qualitative effect. The stronger the coupling the weaker and lesser is 
 the current magnification in any energy scale. As for energies where higher 
 number of modes are propagating and number of cancellations of current 
 magnification is also high we obtained even less current 
 magnifications and their occurence frequency in the energy axis are also 
 reduced (Fig. \ref{result-repul-strong}). 
This effect is entirely due to the superposition of currents from all the 
different channels which is absent in purely 1D system. The non-trivial 
effect of system-reservoir coupling on the equilibrium currents in 1D quantum 
double ring system has been discussed recently in Ref.\cite{colin-jayan}.

We now consider the case of an attractive impurity $\delta$ function 
potential ( $V < 0$). We see in Fig. \ref{result-attr-strong} that  
the amplitude of 
current magnification is lesser in the stronger coupling regime 
($\eps=0.48$) in comparison to Fig. \ref{result-repul-strong}. The magnitude
and positions of the current peaks are very sensitive to the details
of the system parameters and they can not be predicted apriori.
Moreover, the current magnification effect is always absent at the quasi-bound 
state of the negative potential (Fig.~\ref{result-attr-strong}). 
The energies of the quasi-bound states are marked by arrows in  
Fig. \ref{result-attr-strong}. Quasi-bound states are characterised by peak
in the density of states (DOS) and for further 
discussion on quasi-bound state see Ref.~\cite{bag90,swarnali1,swarnali2}. 
The presence of negative delta-function potential enhances DOS near this 
potential. This 
enhanced local DOS at the impurity site reduces the DOS of the propagating 
electrons thereby reducing the current magnification.

In Fig.~\ref{asym-nearbnd1} we have considered a special case and 
plotted the total transport current 
density $dj_T$ and circulating current density $dj_c$ 
in the energy range $4.6E_0 < E < 5.4E_0$. In this energy range at the
Fermi energy there are two propagating modes. The corresponding bound-state
is at $5.1025E_0$. Around the bound-state there is an enhancement in 
scattering. The structure of the total 
transport current exhibits a symmetric line shapes (like Briet-Wigner type
symmetric resonances). Around these resonances we do observe current 
magnification. This special case shows that Fano type resonance structure 
in the total transport 
current is not a necessary criteria for the observation of current 
magnification effect.
      
\section{conclusion}
\label{S4}
In conclusion, we have shown that for a system weakly coupled with reservoirs 
current magnification is a robust effect even in multi-channel case in
the presence of transport current. The magnitude of the circulating current 
can be very large even in presence of several propagating modes despite mode 
mixing and cancellation effects as discussed in the text.
The circulating current are mostly associated with Fano resonance in total 
transport current. However, there are sometimes exception to this rule,
namely, current magnification may occur around Briet-Wigner type
symmetric resonances in the total current. Unlike purely one dimensional system Fano resonance does not
exhibit the zero in the total transmission, however, it is characterised by a 
sharp minimum along with asymmetric line shapes in the total current. 
Impurity strength can enhance or suppress current magnification and is
sensitively dependent on system parameters.
We have established that the system-reservoir coupling strength controls the 
current magnification qualitatively. As the coupling
becomes stronger the current magnification becomes weaker and its occurence
in the given energy range reduces. Thus system reservoir coupling parameter 
controls the transport properties in a very interesting manner. It is interesting to note that persistent currents in a ballistic mesoscopic ring in the 
presence of magnetic flux increases with the Fermi energy (or the number of
channels)~\cite{butt83}. In contrast the magnitude of the current magnification is independent of the total number of propagating channels. It may be 
emphasized that persistent currents and the circulating currents due to current
magnification are two independent distinct phenomena~\cite{pareek}.
\section{acknowledgments}
\label{S5}
One of the authors (S.B.) thanks Prof. Binayak Dutta Roy,
Debasish Chaudhuri and Raishma Krishnan for several useful discussions. 
S.B. also thanks Institute of Physics, Bhubaneswar for providing local 
hospitality where part of the work was carried out.


\end{document}